\documentclass[twocolumn,prb,showpacs,preprintnumbers,amsmath,amssymb,aps,10pt]{revtex4-1}

\usepackage{graphicx}
\usepackage{dcolumn}
\usepackage{bm}
\usepackage{verbatim}

\usepackage[dvips]{color}  
\usepackage{ulem} 

\begin{document}


\title{Comment on ``Diamagnetism and Cooper pairing above $T_c$ in cuprates''}

\author{R.I.~Rey}
\author{A.~Ramos-\'{A}lvarez}
\author{J.~Mosqueira}
\author{M.V.~Ramallo}
\author{F.~Vidal}

\affiliation{LBTS, Facultade de F\'isica, Universidade de Santiago de Compostela, ES-15782 Santiago de Compostela, Spain}

\date{\today}

\begin{abstract}
It is shown that the magnetization rounding measured by L.~Li and coworkers above the superconducting transition in optimally doped YBa$_2$Cu$_3$O$_{7-\delta}$ crystals under magnetic fields up to 14 T [Phys. Rev. B \textbf{81}, 054510 (2010)] may be explained at a phenomenological level in terms of the mean field Gaussian-Ginzburg-Landau (GGL) approach for layered superconductors. This result challenges the claims of L.~Li and coworkers, who write ``\textit{[...] we are observing the phase-disordering mechanism, rather than Gaussian mean-field fluctuations}'', but it is in full agreement with earlier magnetization measurements by different authors in optimally-doped YBa$_2$Cu$_3$O$_{7-\delta}$ under lower magnetic fields. The adequacy of the mean-field Ginzburg-Landau descriptions is further confirmed when analyzing the magnetization data reported below $T_c$ by L.~Li and coworkers.
\end{abstract}

\maketitle

In Ref.~\onlinecite{uno} it is claimed that the magnetization rounding measured in that work above $T_c$ (the so-called precursor diamagnetism) in several families of high temperature cuprate superconductors (HTSC) provides thermodynamic evidence of the so-called phase fluctuation scenario, a conclusion also stressed in a Viewpoint on that paper by Kivelson and Fradkin.\cite{dos} In the case of the prototypical optimally doped (OPT) YBa$_2$Cu$_3$O$_{7-\delta}$ (YBCO), in Ref.~\onlinecite{uno} it is noted that it ``\textit{[...] should be the least susceptible to the phase-disordering mechanism for the destruction of long-range phase coherence at $T_c$ (and hence the best candidate for Gaussian fluctuations among cuprates). However, the torque measurements reveal that $T_c$ in OPT YBCO is also dictated by large phase fluctuations}.'' To support these conclusions, Li and coworkers provide in Fig.~8 of their paper magnetization data up to magnetic fields of 14 T.\cite{uno} Then, without presenting any quantitative comparison with earlier Gaussian-Ginzburg-Landau (GGL) approaches for the precursor diamagnetism in  superconductors,\cite{tres,cuatroAA,cuatro,cinco,seis,siete,ocho,nueve} they write that the ``\textit{[...] significant diamagnetism surviving to intense fields, at temperatures up to 40 K above $T_c$ is strong evidence that we are observing the phase-disordering mechanism, rather than Gaussian mean-field fluctuations}''. 

In principle, the credibility and relevance of the conclusions summarized above could be strongly enhanced by the well-known experimental advantages of the OPT YBCO crystals, which allow a reliable extraction of the precursor diamagnetism.\cite{diez} Also, as stressed in Refs.~\onlinecite{uno} and \onlinecite{dos}, the torque magnetometry technique used in these measurements allows the use of large applied magnetic fields.\cite{once} However, in this Comment we show that, contrary to the claims of Refs.~\onlinecite{uno} and \onlinecite{dos}, these interesting experimental results may be easily explained at a quantitative level in terms of the GGL approach for multilayered superconductors,\cite{cinco,seis,siete}$^,$\cite{nueve} confirming similar conclusions obtained by different authors from measurements under lower applied magnetic fields.\cite{cinco,seis,siete,ocho,nueve} Let us also note already here that, in our opinion, the opportunity of our present Comment is enhanced by the fact that the questionable conclusions of Ref.~\onlinecite{uno} are being cited by not few authors as a strong experimental support of the phase fluctuation scenario for the HTSC,\cite{catorce} at present a still open and debated issue of the physics of these superconductors.\cite{quince}
 
Some of the arguments presented in Ref.~\onlinecite{uno} against the GGL scenario for the precursor diamagnetism in OPT YBCO may be in fact easily refuted without the need of detailed calculations. For instance, when analyzing the magnetization versus temperature curve presented in their Fig.~8(b), these authors write: ``\textit{[...] the curve at 14 T reveals the existence of the large fluctuating diamagnetism associated with the vortex liquid. This point [...] highlights the major difference between the diamagnetism in hole-doped cuprates and low-$T_c$ superconductors. In the latter, increasing $H$ in the fluctuation regime above $T_c$ rapidly squelches the (Gaussian) fluctuation signal altogether}.'' However, as the in-plane coherence length amplitude, $\xi_{ab}(0)$, of OPT YBCO is around 1 nm,\cite{cinco,seis,siete,ocho,nueve} the reduced magnetic field, $h \equiv H/[\phi_0/2\pi\mu_0\xi_{ab}^2(0)]$, corresponding to 14 T is near 0.05 (in this expression $\phi_0$ is the flux quantum and $\mu_0$ the vacuum permeability). By taking a look at Fig.~8.5 of Tinkham's textbook,\cite{tres} one may already conclude that also in the conventional metallic low-$T_c$ superconductors such a relatively weak reduced field has an almost unobservable influence on their precursor diamagnetism, in full agreement with the results of the GGL approach in presence of finite magnetic fields (the so-called Prange regime).\cite{tres,dieciseis} In fact, the results of Soto and coworkers on the precursor diamagnetism in Pb-In alloys, published eight years ago,\cite{diecisiete} provide a direct experimental demonstration that even the precursor diamagnetism onset, that in the zero-field limit is located at temperatures about 1.7$T_c$, is little affected by reduced magnetic fields below 0.1.

The adequacy of the GGL approach for multilayered superconductors to account for the results of Ref.~\onlinecite{uno} on OPT YBCO may be easily confirmed at a quantitative level by just comparing these data with the theoretical expression for the fluctuation-induced magnetization for fields applied perpendicular to the $ab$ layers (denoted $M_d$, as in Ref.~\onlinecite{uno}) proposed in Ref.~\onlinecite{nueve} for layered superconductors under a total-energy cutoff :
 \begin{eqnarray}
 &&M_d(\varepsilon,h)=-f\frac{k_BT}{2\pi\phi_0}\int_{-\pi/s}^{\pi/s}dk_z\left[-\frac{\varepsilon^c}{2h}\psi\left(\frac{1}{2}+\frac{\varepsilon^c}{2h}\right)\right.\nonumber\\
 &&+\frac{\varepsilon+\omega_{k_z}}{2h}\psi\left(\frac{1}{2}+\frac{\varepsilon+\omega_{k_z}}{2h}\right)+\ln\Gamma\left(\frac{1}{2}+\frac{\varepsilon^c}{2h}\right)\nonumber\\
 &&\left.-\ln\Gamma\left(\frac{1}{2}+\frac{\varepsilon+\omega_{k_z}}{2h}\right)+\frac{\varepsilon^c-\varepsilon-\omega_{k_z}}{2h}\right].
 \label{prange}
 \end{eqnarray}
 Here $k_B$ is the Boltzmann constant, $\Gamma$ and $\psi$ are the gamma and, respectively, digamma functions, $\omega_{k_z}=B_{LD}[1-\cos(k_zs)]/2$ is the out-of-plane spectrum of the fluctuations, $B_{LD}=[2\xi_c(0)/s]^2$ is the so-called Lawrence-Doniach parameter, $\xi_c(0)$ is the coherence length amplitude along the crystal $c$-axis, $s=0.59$~nm is the effective superconducting layers periodicity length,\cite{cinco,seis,siete,ocho,nueve,periodicity} $\varepsilon\equiv \ln(T/T_c)$ is the reduced-temperature, $\varepsilon^c\approx 0.5$ is the total-energy cutoff constant,\cite{diecinueve} and $f$ the effective superconducting volume fraction.\cite{volumefraction} Note that if the cutoff is neglected and for low reduced magnetic fields, i.e., for $h,\varepsilon\ll \varepsilon^c$, this expression recovers the well-known expression proposed by Klemm,\cite{seis} which as shown in Ref.~\onlinecite{siete} can be recast in a conventional Lawrence-Doniach\cite{cuatro} form (also proposed independently by Tsuzuki\cite{tsuzuki} and by Yamaji\cite{yamaji}): 
\begin{equation}
M_d(\varepsilon,h)=-f\frac{k_BT}{6\phi_0s}\frac{h}{\varepsilon}\left(1+\frac{B_{LD}}{\varepsilon}\right)^{-1/2}.
\label{schmidt}
\end{equation}

The data points in Fig.~1(a) correspond to those presented in Fig.~8(a) of Ref.~\onlinecite{uno} for the $M_d$ dependence on the applied magnetic field above $T_c$. The solid lines correspond to Eq.~(\ref{prange}) evaluated by using the $T_c$ value proposed in Ref.~\onlinecite{uno} (92~K), the coherence length amplitudes in Ref.~\onlinecite{seis}, $\xi_{ab}(0) = 1.1$~nm and $\xi_c(0) = 0.1$~nm,\cite{coherence} and $f = 0.5$ which is well within the one observed in twinned OPT YBCO crystals.\cite{dieciocho} As it may be seen, the agreement is reasonably good, taking into account the unavoidable uncertainties associated with the determination of both the corresponding background contributions and the $T_c$.\cite{notaTc} The agreement extends also to the as-measured magnetization (denoted $M_{\rm eff}$, as in Ref.~\onlinecite{uno}) versus temperature curve presented in Fig.~8(b) in Ref.~\onlinecite{uno}, which corresponds to an applied magnetic field of 14 T. The analysis of this curve on the grounds of Eq.~(\ref{prange}) is presented in our Fig.~1(b). The solid line was obtained by adding the background proposed in Fig.~8(b) of Ref.~\onlinecite{uno} (doted curve) and the fluctuation-induced magnetization calculated from Eq.~(\ref{prange}) by using the same parameter as before for Fig.~1(a). As one may appreciate, the agreement is excellent even in the high reduced-temperature region, including the temperature for the onset of fluctuation effects. Our present results are also consistent with earlier analyses of Lee, Klemm and Johnston\cite{cinco} and of Ramallo, Torr\'on and Vidal,\cite{siete} although in their case without penetrating the high reduced-temperature region. Surprisingly, in Refs.~\onlinecite{uno} (and \onlinecite{dos}) it was not presented any quantitative analysis of their experimental results on the grounds of the GGL approaches. In addition, earlier results obtained under lower field amplitudes by other authors, as those published in Refs.~\onlinecite{cinco} to \onlinecite{nueve}, were also overlooked. 

%
%

\begin{figure}[t]
\includegraphics[scale=.55]{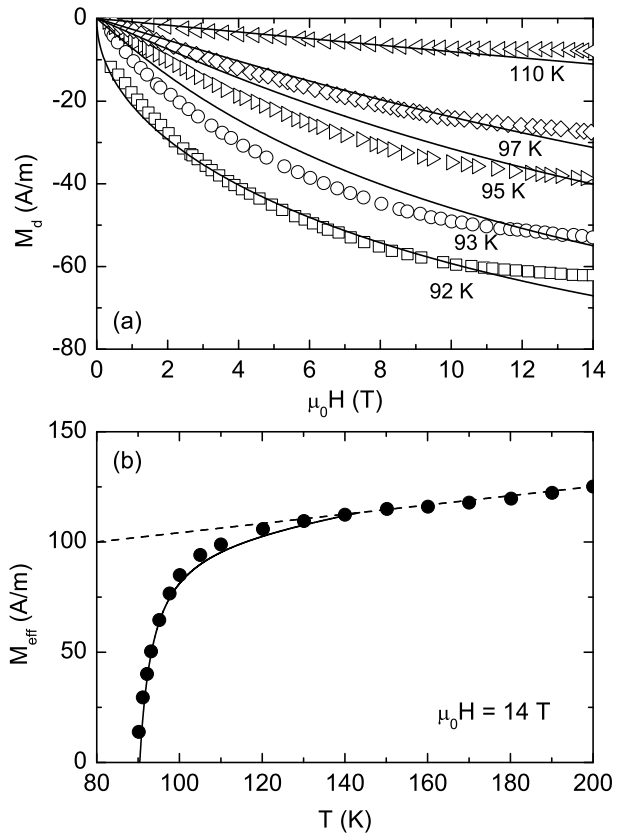}
\caption{The data points in (a) and (b) correspond to the magnetization measurements of Ref.~\onlinecite{uno} in OPT YBCO (Fig.~8) for temperatures above $T_c$. The solid lines in both figures were obtained from the extended GGL approach [Eq.~(\ref{prange})], by using the same coherence length amplitudes as in Ref.~\onlinecite{nueve} when analysing the precursor diamagnetism measured in OPT YBCO under lower applied magnetic fields. The dashed line in (b) is the background contribution. For details see the main text. }
\end{figure}

Although both the experimental results and the, always qualitative, analyses on the diamagnetism in OPT YBCO presented by Li and coworkers\cite{uno} are centered on the behaviour above $T_c$, for completeness we present here a detailed quantitative analysis, in terms of the Ginzburg-Landau scaling proposed by Ullah and  Dorsey,\cite{Ullah} of the data below $T_c$ provided by these authors. This approach corresponds to the lowest-Landau-level (GL-LLL) approximation for 3D superconductors. For applied magnetic fields up to 5 T, the adequacy of this GL-LLL approach to explain the diamagnetism around $T_c$ observed in OPT YBCO was first demonstrated by Welp and coworkers,\cite{Welp} and then by other authors.\cite{dieciocho,Pierson} All these results were also overlooked by Li and coworkers.

%
%

\begin{figure}[t]
\includegraphics[scale=.55]{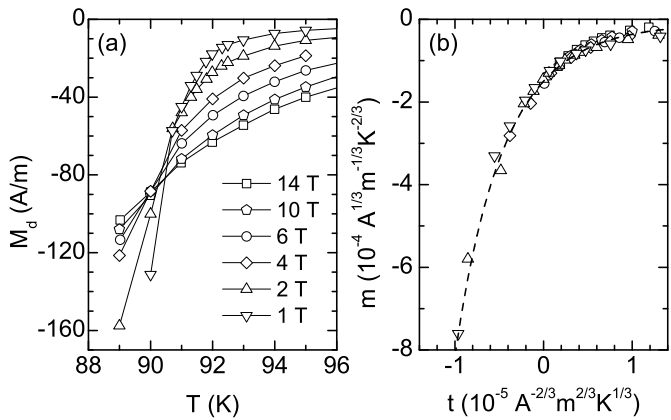}
\caption{(a) Detail of the temperature dependence of $M_d$ around $T_c$ for fields up to 14 T [extracted from Fig.~8(a) in Ref.~\onlinecite{uno}]. (b) 3D GL-LLL scaling of the data in (a). The scaling variable $t$ was evaluated by using the superconducting parameters resulting from the analysis in the Gaussian region (see the main text for details). All the lines are guide for the eyes.}
\end{figure}

The illustrative $M_d(T,H)$ data around $T_c$ that may be extracted from Fig.~8(a) of Ref.~\onlinecite{uno} are represented in our Fig.~2(a). Let us note first that as the slope of the upper critical field, $H_{c2}$, at $T_c$, $\mu_0|dH_{c2}/dT|_{T_c}=\phi_0/2\pi\xi_{ab}^2(0)T_c$, is around 3~T/K, most of these data (down to 89~K and up to 14 T) will fall into the \textit{critical} fluctuation region around the $H_{c2}(T)$ line, where the GL-LLL approach is applicable. Then, on the grounds of this approach $M_d(T,H)$ should follow a scaling behaviour in the variables\cite{Ullah}
 \begin{equation}
 t\equiv\frac{T-T_c(H)}{(TH)^x}
 \label{t}
 \end{equation}
 and 
 \begin{equation}
 m\equiv\frac{M_d}{(TH)^x},
 \label{m}
 \end{equation}
 where $x=2/3$ in the case of three-dimensional (3D) systems. The resulting scaling is shown in Fig.~2(b). In applying the above variables, $T_c(H)$ in Eq.~\ref{t} was obtained from 
 \begin{equation}
 T_c(H)=T_c\left[1-\frac{H}{\phi_0/2\pi\mu_0\xi_{ab}^2(0)}\right],
 \end{equation} 
 by using for $T_c$ and $\xi_{ab}(0)$ the values resulting from the analysis performed in the Gaussian region above $T_c$. Therefore, the present results not only nicely extend the applicability of GL-LLL approaches in OPT YBCO in the critical region around $T_c(H)$ to fields as large as 14~T, but they also represent a stringent check of consistency with the GGL analyses we have presented here for the data of Ref.~\onlinecite{uno} above $T_c$. It is also worth noting that the consistency between GL approaches for the fluctuation effects above and below $T_c$ was also shown for highly anisotropic HTSC.\cite{consistency}

 Let us note also that the seemingly anomalous non-monotonic profile of the $M_d$ vs. $H$ isotherms that may be observed below $T_c$ in Fig.~8(a) of Ref.~\onlinecite{uno} is compatible with the conventional (GL-like) fluctuation scenario: in YBCO the slope of the upper critical field close to $T_c$ is $\mu_0|dH_{c2}/dT|_{T_c}\approx3$~T/K. Then, $\mu_0H_{c2}$ just below $T_c$ is in the Tesla range. In particular, at 91, 90, and 89 K it is about 3, 6, and 9 T respectively. A simple inspection of the corresponding $M_d(H)$ reveals a change from concave to convex just at these field values. This  change may be then attributed to a transition from the mixed state (in which $|M_d|$ decreases with $H$) to the normal state at higher fields (in which, due to conventional fluctuation effects, $M_d$ is not null and behaves like in the isotherms above $T_c$).
  
 We must also remark here that the differences between the diamagnetism behavior in the OPT YBCO and in a low-$T_c$ superconductor may be easily attributed to the differences between their superconducting parameters. For instance, in the case of the NbSe$_2$ compared in Ref.~\onlinecite{Wang} with the HTSC, $T_c$ is one order of magnitude smaller than in OPT YBCO, while the coherence lengths are one order of magnitude larger. Then, for similar reduced temperatures and magnetic fields, the fluctuation magnetization above $T_c$ is a factor $\sim$300 smaller as may be easily estimated from the GGL approach. A thorough comparison between the diamagnetism above $T_c$ in HTSC and low-$T_c$ superconductors, and on the corresponding influence of the presence of $T_c$ inhomogeneities, is presented in Ref.~\onlinecite{veinte} and in the references therein.

 In summary, we have analyzed at a quantitative level, in terms of the mean-field GGL approach for multilayered superconductors, the experimental results of Li and coworkers\cite{uno} on the diamagnetism above $T_c$ in OPT YBCO. These analyses lead to conclusions just opposite to those stressed in Ref.~\onlinecite{uno} (and Ref.~\onlinecite{dos}), but provide a nice quantitative confirmation, up to magnetic field amplitudes of 14 T, of earlier conclusions of different authors,\cite{cinco,seis,siete,ocho} later extended to high-reduced temperatures through the introduction of a total-energy cutoff:\cite{nueve,diecinueve} the adequacy of the extended GGL scenario to account at a phenomenological level for the precursor diamagnetism (including its onset) of this prototypical cuprate superconductor. Although not addressed in Ref.~\onlinecite{uno}, for completeness we have also analysed quantitatively, in terms of the Ginzburg-Landau scaling proposed by Ullah and Dorsey,\cite{Ullah} the data below $T_c$ provided in that work. This comparison demonstrates the adequacy of the GL scaling for the data located in the critical region around $T_c(H)$, extending  up to 14~T the earlier conclusions for lower field amplitudes.\cite{dieciocho,Welp,Pierson}  As these analyses above and below $T_c$ have been performed by using the same parameter-values, our results represent a stringent check of consistency of the validity of the mean-field-like GL scenario for the fluctuation diamagnetism observed above and below the superconducting transition in OPT YBCO.  It would be useful to extend these analyses to the results of Ref.~\onlinecite{uno} on other HTSC families with different doping levels. However, as we have already stressed when commenting other results of these authors,\cite{veinte} non-optimally-doped compounds will be more affected by chemical disorder (intrinsic to their non-stoichiometric nature)\cite{veintiuno} than OPT YBCO, and the superconducting fluctuation effects will be entangled with those associated with the corresponding $T_c$-inhomogeneities.\cite{veinte,veintiuno,veintidos}

 Finally, let us stress that the results summarized here try to contribute, by presenting quantitative analyses on the grounds of the mean-field-like GL approaches, to clarify the debate on the phenomenological descriptions of the superconducting transition in HTSC, but do not pretend to close that debate even in the case of the prototypical OPT YBCO. In fact, the majority of the theoretical works published in the last few years on that issue, including some of the most influential, propose different unconventional (non-Ginzburg-Landau) scenarios, the most popular being the one based on phase-disordering.\cite{dos,quince,oganesyan,podolsky,benfatto,veintitres,veinticuatro,veinticinco,veintisiete,tsevelik,veintinueve, treinta,treintayuno,new1,new2,new3,new4} Even in the case of the conventional GL-like scenarios, the possible presence of different, direct or indirect, superconducting fluctuation contributions is a long standing issue,\cite{siete,ocho} still  open at present.\cite{new5} However, our present Comment further stresses that, in any case, before adopting any superconducting fluctuation scenario for the HTSC  it will be crucial to perform thorough quantitative analysis of the experimental results, in some cases taking into account the presence of chemical disorder.\cite{veinte,veintiuno,veintidos,new6}

 Supported by the Spanish MICINN and ERDF \mbox{(No.~FIS2010-19807)}, and by the Xunta de Galicia (Nos.~2010/XA043 and 10TMT206012PR). N.C. and A. R-A. acknowledge financial support from Spain's MICINN through a FPI grant.


\begin{references}
\bibitem{notaTc}These uncertainties may justify the slight disagreement observed in the isotherms close to $T_c$.
\end{references}
\end{document}